\documentclass[twocolumn,showpacs,preprintnumbers,amsmath,amssymb, superscriptaddress, prb]{revtex4-1}

\usepackage{stmaryrd}
\usepackage{txfonts}
\usepackage{amssymb}
\usepackage{mathrsfs}
\usepackage{graphicx}
\usepackage{dcolumn}
\usepackage{bm}
\usepackage{epsfig}
\usepackage{color}                    

\begin{document}

\title{Synchronization of Josephson oscillations in mesa array of $\rm{Bi_2Sr_2CaCu_2O_{8+\delta}}$ single crystal through the Josephson
plasma waves in base crystal}
\author{Shi-Zeng Lin}
\email{szl@lanl.gov} \affiliation{Theoretical Division, Los Alamos National
Laboratory, Los Alamos, New Mexico 87545, USA}

\author{Alexei E. Koshelev}
\email{koshelev@anl.gov} \affiliation{Materials Science Division, Argonne
National Laboratory, Argonne, Illinois 60439, USA}

\begin{abstract}
Using mesa array of $\rm{Bi_{2}Sr_{2}CaCu_{2}O_{8}}$ single crystal
was demonstrated recently as a promising route to enhance the radiation power generated by Josephson oscillations in mesas.
We study the synchronization in such an array via the plasma waves
in the base crystal. First, we analyze plasma
oscillations inside the base crystal generated by the synchronized mesa array
and the associated dissipation. We then solve the dynamic equation for
superconducting phase numerically to find conditions for synchronization and to
check the stability of synchronized state. We find that mesas are synchronized
when the cavity resonance of mesas matches with that of the base crystal. An
optimal configuration of mesa arrays is also obtained.
\end{abstract}
\pacs{74.50.+r, 74.25.Gz, 85.25.Cp}
\date{\today}
\maketitle

\section{Introduction}
Soon after the discovery of Josephson Effects, it was realized that Josephson
junction can be used to generate electromagnetic waves. When the junction is
biased in voltage state with voltage $V$, the two superconducting electrodes
have energy difference $2eV$. The system is similar to two-energy level system
in atomic physics. When Cooper pairs tunnel from the electrode with higher
energy to that with lower energy, a photon with angular frequency
$\omega=2eV/\hbar$ is emitted. The frequency can be tuned by voltage and $1$ mV
corresponds to $0.483$ THz. The radiation power from one junction however is
weak, of the order of 1 pW. \cite{Yanson65,Dayem66,Zimmerma66} Arrays of Josephson
junctions are fabricated to enhance the radiation
power\cite{Finnegan72,Jain84,Durala99,Barbara99,Song09}. Once these junctions
are synchronized, the total radiation power is proportional to the number of
junctions squared.

A stack of Josephson junctions is naturally realized in some layered cuprate
superconductor\cite{Kleiner92}, such as $\rm{Bi_2Sr_2CaCu_2O_{8+\delta}}$
(BSCCO). Because of the large superconducting energy gap (60 meV), these
build-in intrinsic Josephson junctions (IJJs) may have Josephson oscillations
with frequencies in the terahertz (THz) band. IJJs are packed on nanometer
scale, much smaller than THz electromagnetic (EM) wavelength, and are
homogeneous. The THz generator based on IJJs thus is promising to fill the THz
gap\cite{Hu10,Savelev10}. Lots of effort has been made to excite the coherent
THz radiation experimentally in the last decade
\cite{Iguchi00b,Batov06,Bae07,Benseman11}. On the theoretical side, numerical
simulations and analytical calculations are performed to understand the
mechanism of radiation.
\cite{Tachiki94,Koyama95,Tachiki05,Bulaevskii06,Bulaevskii07,szlin08,Koshelev08,Tachiki09,szlin09a}

Coherent radiations from a mesa structure of BSCCO in the absence external
magnetic fields were observed experimentally in 2007\cite{Ozyuzer07}, which
renewed the interest in this field. It was found that the mesa itself forms a
cavity to synchronize the radiation in different layers, as evidenced from the
dependence of the radiation frequency $f$ on the lateral size $L$ of the
mesa, $f=c_0/(2L)$ with $c_0=c/\sqrt{\epsilon_c}$ the Josephson plasma
velocity where $\epsilon_c$ is the dielectric constant of BSCCO. The cavity
resonance mechanism has been confirmed by many independent
experiments\cite{kadowaki08,Wang09,Wang10,Tsujimoto10,Tsujimoto12b} and the
radiation power is enhanced to about $30\ \rm{\mu W}$.\cite{Yamaki11} A dynamic
state with $\pi$ phase kink was proposed to account for the experimental
observations \cite{szlin08b,Koshelev08b}. It was suggested that the strong
in-plane dissipation is responsible for the excitation of cavity mode uniform
along the $c$-axis. \cite{szlin12a}

\begin{figure}[b]
\psfig{figure=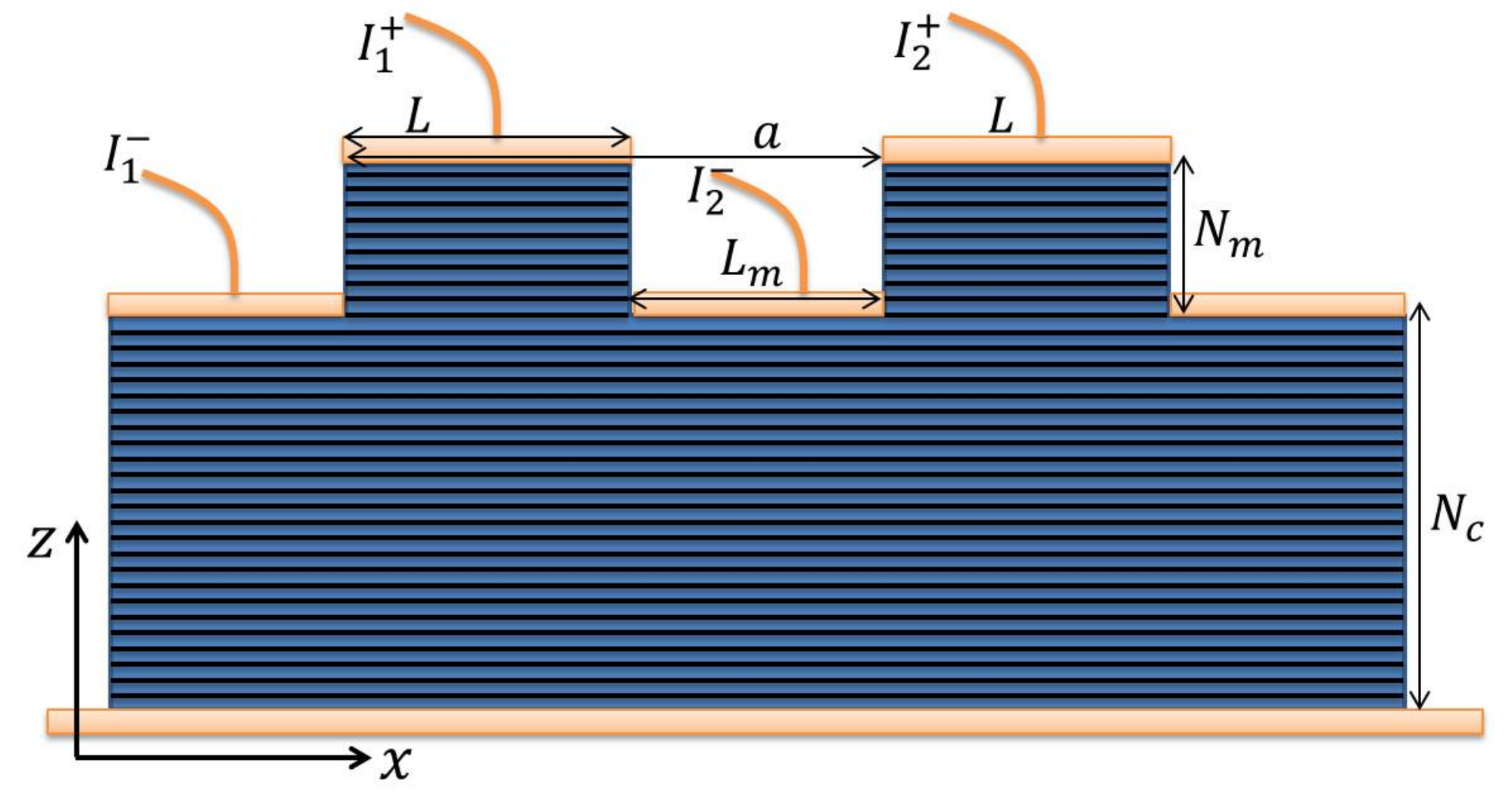,width=\columnwidth}
\caption{\label{f1}(color online)
Schematic view of multiple mesas atop of BSCCO crystal. The mesas are biased independently. The BSCCO sample (blue) is sandwiched by gold electrodes (orange). In analytical treatment, we consider an array of identical mesas with width $L$ and period $a$. The top surface of the base crystal is free. In simulations, we consider two identical mesas with width $L$ separated by a distance $L_m$. The mesas are located at the position $L$ away from the edges of the base crystal. The top surface of the base crystal is also covered by gold electrodes through which the current is extracted.}
\end{figure}

From application perspective, the radiation power in the present experimental
design is still too weak to be practically useful. A natural way to enhance
the radiation power by using thicker mesas has several challenges. First, for a thick mesa it becomes difficult to cool the system efficiently.
The dissipation hence self-heating increases with the volume of the mesa, while
the heat removal rate remains the same because the heat is mainly removed
through the substrate. It has already shown experimentally that even for a mesa
with thickness of $\sim 2\rm{\mu m}$, central part of the meas is driven to the
normal state by the severe self-heating.\cite{Wang09,Wang10}. Secondly, it was calculated that for a tall mesa a long-range instability destroying the in-phase plasma oscillations develops\cite{Koshelev10}, and only parts of the mesa can be synchronized.\cite{szlin12a}

To enhance the radiation power while minimizing the self-heating, one may use
multiple thin mesas on top of the same BSCCO single crystal. The multiple mesa
structure has been fabricated recently and the radiation power is enhanced
under appropriate conditions as demonstrated in the recent
experiments.\cite{Orita2010,Benseman12} The mechanism of synchronization among
mesas is not known. There are two sources of interaction. The mesas interact
through the radiation fields. They also interact through the plasma
oscillations in the base crystal. The resonance damping due to the leaking of radiation from the mesa into the base crystal has been considered in Ref. \onlinecite{Koshelev09} and it was
demonstrated that this channel gives the main contribution to the dissipation. Therefore,
synchronization mediated by radiation fields inside crystal is probably a
dominating mechanism. The present work is devoted to understanding the synchronization of multiple
mesas through plasma oscillations in the base crystal and to finding an optimal
configuration for synchronization.

\section{Model}

We consider arrays of identical mesas with the period $a$ atop of BSCCO single crystal as schematically shown in Fig.
\ref{f1}. Every mesa contains $N_{\mathrm{m}}$ junctions and has
width $L$ while the base crystal contains $N_{c}$ junctions. No external magnetic field is applied to the BSCCO. Each mesa is biased independent by a dc current injected from top of the mesa and extracted from the sides of the mesa. In this case, the junctions in the basal crystal remain zero-voltage state, and the mesas are driven into resistive state. The system is assumed to be uniform along the $y$ direction and the problem becomes two dimensional.
The dynamics of the gauge invariant phase difference $\theta_n$ and magnetic
field $h_{n}$ in the $n$-th junction are described by
\cite{Sakai93,Bulaevskii94,Bulaevskii96,Machida99,Koshelev01}
\begin{equation}\label{eq1}
\frac{\partial^{2}\theta_{n}}{\partial\tau^{2}}+\nu_{c}\frac{\partial
\theta_{n}}{\partial\tau}+\sin\theta_{n}-\ell^{2}\frac{\partial h_{n}
}{\partial x}   =0,
\end{equation}
\begin{equation}\label{eq2}
\left(  \ell^{2}\nabla_{n}^{2}-1\right)  h_{n}+\frac{\partial\theta_{n}
}{\partial x}+\nu_{ab}\frac{\partial}{\partial\tau}\left(  \frac
{\partial\theta_{n}}{\partial x}-h_{n}\right)    =0,
\end{equation}
where $\Delta^{(2)}f_l \equiv f_{l+1}+f_{l-1}-2 f_l$ is the finite difference
operator. Here the time and coordinate are measured in units of the inverse Josephson plasma
frequency $1/\omega_{p}$ and the Josephson length $\lambda_{J}=\gamma s$
correspondingly and the unit of magnetic field is $\Phi_{0}/(2\pi\gamma
s^{2})$, where $\gamma$ is the anisotropy factor and $s$ is the interlayer
spacing. Here $\omega_p=c/(\lambda_c\sqrt{\epsilon_c})$ and $\Phi_0=hc/(2e)$ is the flux quantum. These reduced equations depend on three parameters, $\nu
_{c}\!=\!4\pi\sigma_{c}/(\varepsilon_{c}\omega_{p})$, $\nu_{ab}\!=\!4\pi
\sigma_{ab}/(\varepsilon_{c}\omega_{p}\gamma^{2})$, and $\ell\!=\!\lambda_{ab}/s$,
where $\sigma_{c}$ and
$\sigma_{ab}$ are the quasiparticle conductivity, and $\lambda_{c}$ and $\lambda_{ab}$ are the London penetration depth along the $c$-axis and $ab$-plane respectively. The dimensionless electric field is given by $E_{z,n}=\partial_t\theta_n$, with $E$ in unit of $\Phi_0\omega_p/(2\pi c s)$.

For the mesa with thickness of $1\ \rm{\mu m}$ which is much smaller than the
wave length of THz EM wave in vacuum, there is a significant impedance mismatch
between the mesa and vacuum\cite{Bulaevskii06PRL}. Most part of energy is
reflected at the edges of mesa and cavity resonance is achieved. We can use the
boundary condition that the oscillating magnetic field vanishes at the edges.
The boundary condition at the edges of the mesas is $\partial_x \theta_n=\pm
L I_{\mu}/2$, and the boundary condition at the edges of the base crystal is
$\partial_x \theta_n=0$, where $I_{\mu}$ is the bias
current in the $\mu$-th mesa. We assume that the IJJs stack is sandwiched by good
conductors, such that the tangential current inside the conductor is zero,
which corresponds to the boundary condition $\partial_z h(z)=0$ in the
continuum limit.

\section{Plasma oscillations and associated dissipation in the synchronized state}

In this section, we calculate the plasma oscillations and its dissipation in the synchronized state assuming that the array contains large number of mesas so that it can be treated as an infinite system.
The time dependence of the phases in mesas in resistive state has the form $\theta
_{n}(x,t)=\omega\tau+\varphi_{n}(x)+\operatorname{Re}[\theta_{n}
(x)\exp(-i\omega\tau)]+\psi_{\mu}$ and in the crystal the phases have only
have small oscillations. Here $\psi_{\mu}$ accounts for the phase shift between synchronized mesas. We consider the case with $\psi_{\mu}=0$ and leave the more general case for numerical simulation in the next Section. We consider voltage range corresponding to the
Josephson frequency close to fundamental cavity resonance
$\omega\approx\omega_{1} =\ell\pi/L$. For definiteness, we assume that in mesas
the kink state is formed \cite{szlin08b,Koshelev08b} providing strong coupling
to the cavity resonance meaning that we can use approximations $\varphi_{n}
(x)\approx\pi\mathrm{sgn}(x)$ and $\sin\theta_{n}=\operatorname{Re}
[i\exp(-i\omega\tau-i\varphi_{n}(x))]\approx g(x)\operatorname{Re}
[i\exp(-i\omega\tau)]$ with $g(x)\approx\mathrm{sgn}(x)$. However, a particular
shape of the modulation function $g(x)$ has no importance in further
derivations. For isolated mesa on the top of bulk crystal this problem was
considered in Ref.\ \onlinecite{Koshelev09} where it was concluded that leaking
radiation into crystal provides dominating mechanism of resonance damping.

The amplitudes of phases and magnetic fields obey the following
equations: in mesas for $|x-ma|<L/2$, $0<n<N_{\mathrm{m}},$
\begin{align}
\left(  \omega^{2}+i\nu_{c}\omega\right)  \theta_{n}+\ell^{2}\frac{\partial
h_{n}}{\partial x}  &  =ig(x),\label{EqMesaPh}\\
\ell^{2}\nabla_{n}^{2}h_{n}-\left(  1-i\nu_{ab}\omega\right)  h_{n}+\left(
1-i\nu_{ab}\omega\right)  \frac{\partial\theta_{n}}{\partial x}  &  =0,
\label{EqMesaH}
\end{align}
and in the crystal, for $-N_{c}<n\leq0$, the first equation has to be modified as
\begin{equation}
\left(  \omega^{2}-1+i\nu_{c}\omega\right)  \theta_{n}+\ell^{2}\frac{\partial
h_{n}}{\partial x}=0. \label{EqCrystPh}
\end{equation}

\begin{figure}[t]
\psfig{figure=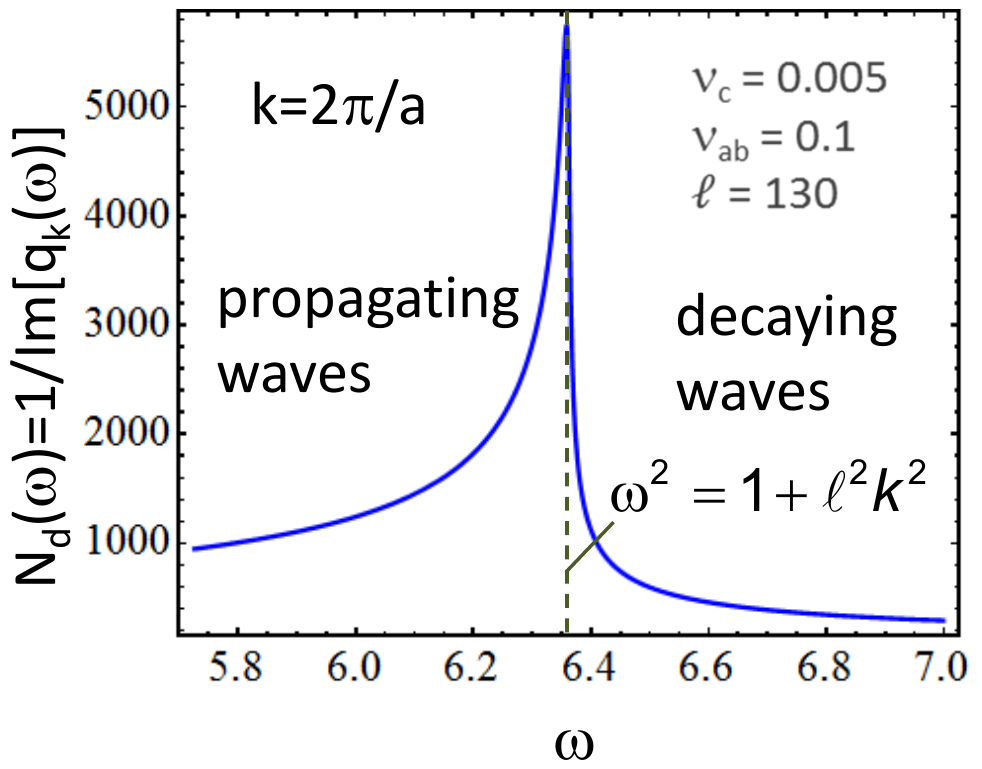,width=0.9\columnwidth}\caption{(color online)
The frequency dependence of the decay length along $c$-direction, $N_d$, for the fixed in-plane
wave vector and representative parameters specified in the plot.}
\label{f2}
\end{figure}

Using presentation $g(x)=\sum_{m=0}^{\infty}g_{m}\sin\left(  p_{m}x\right)$
with $p_{m}=(2m-1)\pi/L$, we can find solution of these equations as mode
expansions. In particular, the oscillating magnetic field in mesas can be
written as
\begin{equation}
h_{n}^{(\mathrm{m})}(x)=\sum_{m=1}^{\infty}\frac{ip_{m}(g_{m}\!+\!a_{m}
\cos\left[  q_{m}\left(  n\!-\!N_{\mathrm{m}}\!-\!1/2\right)  \right]  )}{\omega^{2}+i\nu_{c}
\omega-\ell^{2}p_{m}^{2}}\cos\left(  p_{m}x\right)  \label{hn-mesa}
\end{equation}
for $n>0$, where $q_{m}\equiv q(p_{m},\omega)$ with $\mathrm{Im}(q_m)>0$ is the
wave vector describing propagation of the plasma wave along $c$-axis for the
fixed in-plane wave vector
and frequency,
\begin{equation}
\cos q_{m}=1+\frac{1-i\nu_{ab}\omega}{2\ell^{2}}\left(  1-\frac{\ell^{2}
p_{m}^{2}}{\omega^{2}+i\nu_{c}\omega}\right)
\label{qm}
\end{equation}
In crystal, $n\leq0$, the oscillating magnetic field can be presented as
Fourier series,
\begin{align}
h_{n}^{(\mathrm{cr})}(x)  &  =\frac{1}{a}\sum_{k=2\pi l/a}H_{k}\cos\left[
q_{k}\left(  n+N_{c}+1/2\right)  \right]  \exp(ikx)\label{hn-cryst}\\
\text{with }  &  \cos q_{k}=1+\frac{1-i\nu_{ab}\omega}{2\ell^{2}}\left(
1-\frac{\ell^{2}k^{2}}{\omega^{2}-1+i\nu_{c}\omega}\right)  .\nonumber
\end{align}
It is crucial that the synchronized mesa array excites discrete set of modes
inside the crystal. For the fixed $k$ the frequency range $\omega
^{2}<1+\ell^{2}k^{2}$ corresponds to propagating waves along the $c$-axis while
the range $\omega^{2}>1+\ell^{2}k^{2}$ corresponds to evanescent waves. At the
frequency $\omega^{2}=1+\ell^{2}k^{2}$ the uniform plasma mode is excited. The
decay length of the plasma mode in terms of the number of junctions, $N_{d}(\omega)=1/\mathrm{Im}(q_{k}(\omega))$,
has a sharp maximum at this frequency, see Fig.\ \ref{f2}. For
$a\approx 2L$ the mode with the wave vector $k=2\pi/a$ plays the most important
role, because frequency of the uniform mode
$\omega=\sqrt{1+\ell^{2}(2\pi/a)^{2}}$ is close to the cavity-resonance
frequency inside the mesa $\omega_1=\ell \pi/L$.

The unknown coefficients $a_{m}$ and $H_{k}$ have to be found from matching at
the interface, $h_{n}^{(\mathrm{m})}(x)=h_{n}^{(\mathrm{cr})}(x)$ for $n=0,1$.
Taking the projection of the equation $h_{0}^{(\mathrm{m}
)}(x)=h_{0}^{(\mathrm{cr})}(x)$ to mode $m$, using
\begin{equation}
S_{m}(k)=\int_{-L/2}^{L/2}dx\cos\left(  p_{m}x\right)  \cos kx=\frac
{2(-1)^{m}p_{m}\cos\left[  kL/2\right]  }{p_{m}^{2}-k^{2}},
\end{equation}
we obtain equation expressing $a_{m}$ via $H_{k}$
\begin{align*}
& \frac{ip_{m}\left(  g_{m}+a_{m}\cos\left[  q_{m}\left(  N_{\mathrm{m}}\!+\!1/2\right)
\right]  \right)  }{\omega^{2}+i\nu_{c}\omega-\ell^{2}p_{m}^{2}}\\
& =\frac{1}{a}\sum_{k=2\pi l/a}H_{k}\cos\left[  q_{k}\left(  N_{c}+1/2\right)
\right]  \frac{4(-1)^{m}}{L}\frac{p_{m}\cos\left[  kL/2\right]  }{p_{m}
^{2}-k^{2}}.
\end{align*}
Note that $S_{m}(k)$ satisfy orthogonality conditions
\[
\frac{2}{aL}\sum_{k=2\pi l/a}S_{m}(k)S_{m^{\prime}}(k)=\delta_{mm^{\prime}}.
\]
On the other hand, the inverse Fourier transform of the equation
$h_{1}^{(\mathrm{m})}(x)=h_{1}^{(\mathrm{cr})}(x)$ allows us to express
$H_{k}$ via $a_{m}$
\begin{align}
H_{k}\cos\left[  q_{k}\left(  N_{c}+3/2\right)  \right]    & =\sum
_{m=1}^{\infty}\frac{ip_{m}\left(  g_{m}+a_{m}\cos\left[  q_{m^{\prime}
}\left(  N_{\mathrm{m}}-1/2\right)  \right]  \right)  }{\omega^{2}+i\nu_{c}\omega-\ell
^{2}p_{m}^{2}}\nonumber\\
& \times\frac{2(-1)^{m}p_{m}\cos\left[  kL/2\right]  }{p_{m}^{2}-k^{2}}
\label{Hkcryst}
\end{align}
Eliminating $H_{k}$, we obtain the linear equations for $a_{m}$
\begin{align}
& \left(  \cos\left[  q_{m}\left(  N_{\mathrm{m}}-1/2\right)  \right]  -\cos\left[
q_{m}\left(  N_{\mathrm{m}}+1/2\right)  \right]  \right)  a_{m}\nonumber\\
& -\sum_{m^{\prime}=1}^{\infty}\mathcal{J}_{mm^{\prime}}a_{m^{\prime}}
\cos\left[  q_{m^{\prime}}\left(  N_{\mathrm{m}}-1/2\right)  \right]  =\sum_{m^{\prime}
=1}^{\infty}\mathcal{J}_{mm^{\prime}}g_{m^{\prime}}
\label{am-Eq}
\end{align}
with the matrix
\begin{align*}
\mathcal{J}_{mm^{\prime}}  & =\frac{p_{m^{\prime}}\left(  \omega^{2}+i\nu
_{c}\omega-\ell^{2}p_{m}^{2}\right)  }{p_{m}\left(  \omega^{2}+i\nu_{c}
\omega-\ell^{2}p_{m^{\prime}}^{2}\right)  }\frac{2}{aL}\\
& \times\sum_{k=2\pi l/a}S_{m}(k)S_{m^{\prime}}(k)\left(  1-\frac{\cos\left[
q_{k}\left(  N_{c}+1/2\right)  \right]  }{\cos\left[  q_{k}\left(
N_{c}+3/2\right)  \right]  }\right).
\end{align*}

\begin{figure}[t]
\psfig{figure=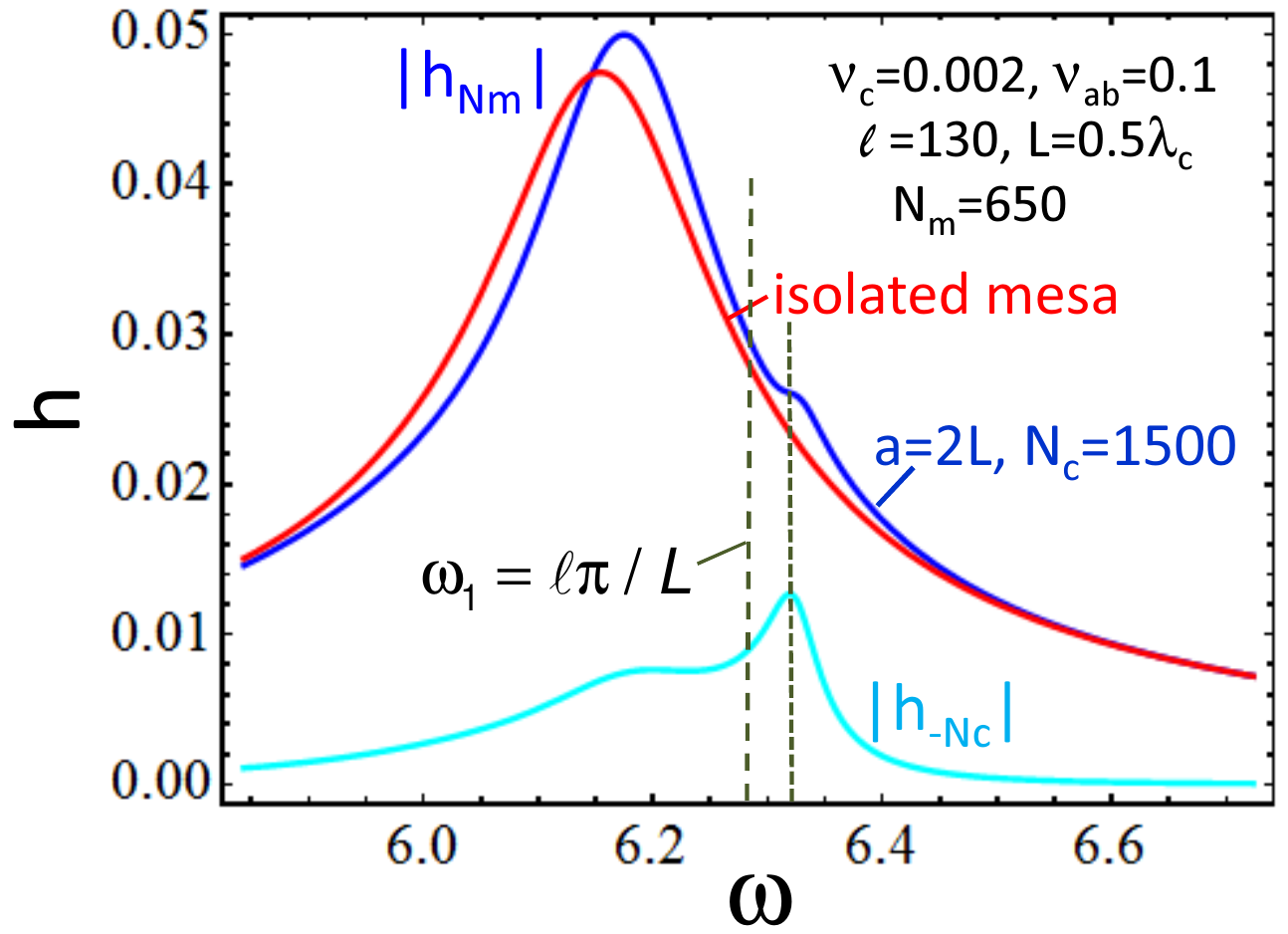,width=0.9\columnwidth}\caption{(color online) Comparison of
the frequency dependence of the oscillating magnetic field at the mesa top
for isolated mesa and mesa array. The frequency dependence of the
oscillating magnetic field at the bottom of the base crystal for mesa array is also shown.
Parameters used in calculations are specified in the plot.}
\label{f3}
\end{figure}
\begin{figure}[b]
\psfig{figure=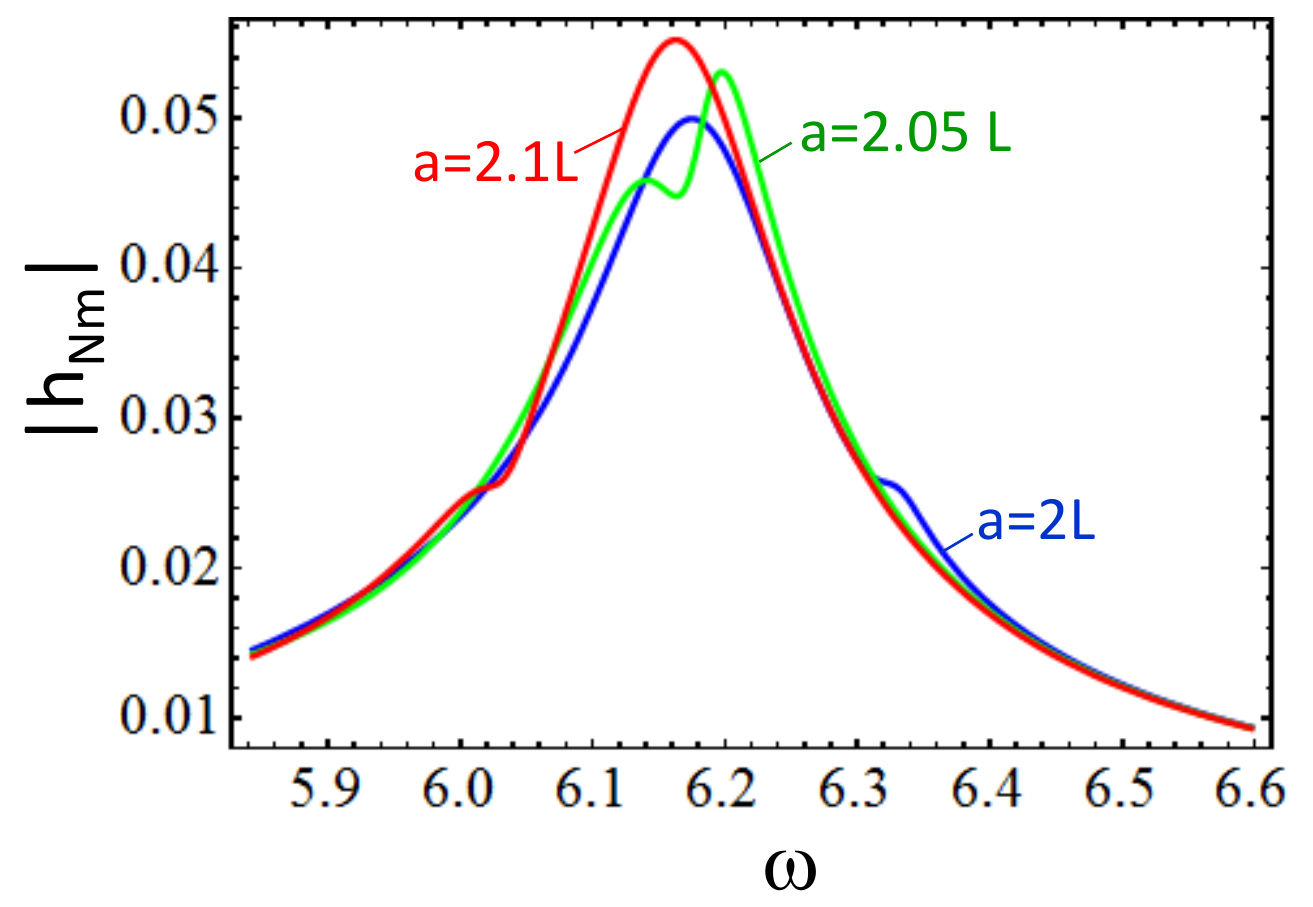,width=0.9\columnwidth}\caption{(color online) The frequency
dependences of the oscillating magnetic field at the mesa top for different
periods of the mesa array. Other parameters are the same as in the previous plot.}
\label{f4}
\end{figure}

Near the fundamental-mode resonance $\omega\approx\ell p_{1}$ the amplitude
$a_{1}$ dominates and we can use single-mode approximation neglecting all other
amplitudes. This leads to a simple result
\begin{align}
a_{1}  &  \approx\frac{g_{1}\mathcal{J}_{11}}{\cos\left[  q_{1}\left(
N_{\mathrm{m}}-1/2\right)  \right]  -\left(  1+\mathcal{J}_{11}\right)  \cos\left[
q_{1}\left(  N_{\mathrm{m}}+1/2\right)  \right]  }\nonumber\\
&  \approx\frac{g_{1}\mathcal{J}_{11}}{q_{1}\sin\left[  q_{1}N_{\mathrm{m}}\right]
-\mathcal{J}_{11}\cos\left[  q_{1}\left(  N_{\mathrm{m}}+1/2\right)  \right]  }\label{a0-approx}
\end{align}
with
\begin{equation}
\mathcal{J}_{11}=\frac{2}{aL}\sum_{k=2\pi l/a}\frac{4p_{1}^{2}\cos^{2}\left[
kL/2\right]  }{(p_{1}^{2}-k^{2})^{2}}\left(  1-\frac{\cos\left[  q_{k}\left(
N_{c}+1/2\right)  \right]  }{\cos\left[  q_{k}\left(  N_{c}+3/2\right)
\right]  }\right)  \label{J00}
\end{equation}
and $q_{1}\approx\sqrt{-\frac{1-i\nu_{ab}\omega}{\ell^{2}}\frac{\omega
^{2}-\ell^{2}p_{1}^{2}+i\nu_{c}\omega}{\omega^{2}+i\nu_{c}\omega}}$. With this
result, one can obtain the oscillating phases and fields in the mesa. Also
using Eq. (\ref{Hkcryst}) and keeping only $m=1$ in the sum, we obtain the
coefficients $H_{k}$ which determine the oscillating magnetic field inside the
crystal, Eq. (\ref{hn-cryst}). For an isolated mesa on bulk crystal
corresponding to the limit $a,N_{c}\rightarrow+\infty$, the following
approximate result can be derived\cite{Koshelev09} $\mathcal{J}_{11}\approx{\sqrt{1-i\nu_{ab}\omega}}\left(0.57+0.31i\right)/{\ell}$, suggesting the following presentation $
\mathcal{J}_{11}={\sqrt{1-i\nu_{ab}\omega}}\beta_{1}/{\ell}$,
where $\beta_{1}$ is the complex function of the order unity. The amplitude of
the oscillating magnetic field on the top of the mesa can be represented
as
\begin{equation}
h_{N_{\mathrm{m}}}^{(\mathrm{m})}(x)\approx\frac{ip_{1}g_{1}\cos\left(  p_{1}x\right)
}{\omega^{2}-\ell^{2}p_{1}^{2}+i\nu_{c}\omega+\mathcal{A}_{1}(\omega)},
\label{hmesatop}
\end{equation}
where the complex function $\mathcal{A}_{1}(\omega)$
\[
\mathcal{A}_{1}\!=\!-\frac{\left(  \omega^{2}-\ell^{2}p_{1}^{2}
+i\nu_{c}\omega\right)  \beta_{1}}{\sqrt{-\frac{\omega^{2}-\ell^{2}p_{1}
^{2}+i\nu_{c}\omega}{\omega^{2}+i\nu_{c}\omega}}\sin\left[  q_{1}N_{\mathrm{m}}\right]
\!+\!\beta_{1}\!\left(  1\!-\!\cos\left[q_{1}\!\left(  N_{\mathrm{m}}\!-\!\frac{1}{2}\right)
\right]  \right)  }.
\]
determines the damping of the resonance and its frequency shift in all quantities.

Figure \ref{f3} illustrates the Josephson-frequency dependence of the
oscillating magnetic field amplitude inside the mesa near the cavity-resonance
frequency for isolated mesa and for mesa array with $a=2L$. One can see that
for used parameters the corrections are weak. Above the main peak one can see a
small dip caused to excitation of the almost uniform standing wave inside the
crystal. To verify this, we also show the plot of the oscillating magnetic
field at the bottom of the base crystal. The dip in the mesa field corresponds
to the rather sharp peak of the field at the bottom of the crystal. Note also
that the resonance is displaced to lower frequency with respect to the uniform
cavity mode $\ell \pi/ L$ because plasma oscillations excited inside the mesa
are not uniform in the c direction. The width of resonance is mostly determined
my leak of radiation into the base crystal.

Figure \ref{f4} shows evolution of the resonance shape with variation of the
array period $a$. We can see that with increasing $a$ the dip moves to smaller
frequencies. The dip has maximum amplitude and located at the peak for
$a=2.05L$. It is interesting to note that the resonance in mesa is actually
strongest when the dip is located \emph{below} the peak at $a=2.1 L$. The
reason is that in this case the wave excited at the peak frequency and the wave
vector $k=2\pi/a$ is in decaying range, see Fig. \ref{f2} and, as consequence,
the mesas loose less energy to radiation at the resonance frequency.
Nevertheless, we expect the strongest interaction between the mesas and optimal
conditions for synchronization when resonances coincide.

\begin{figure}[b]
\psfig{figure=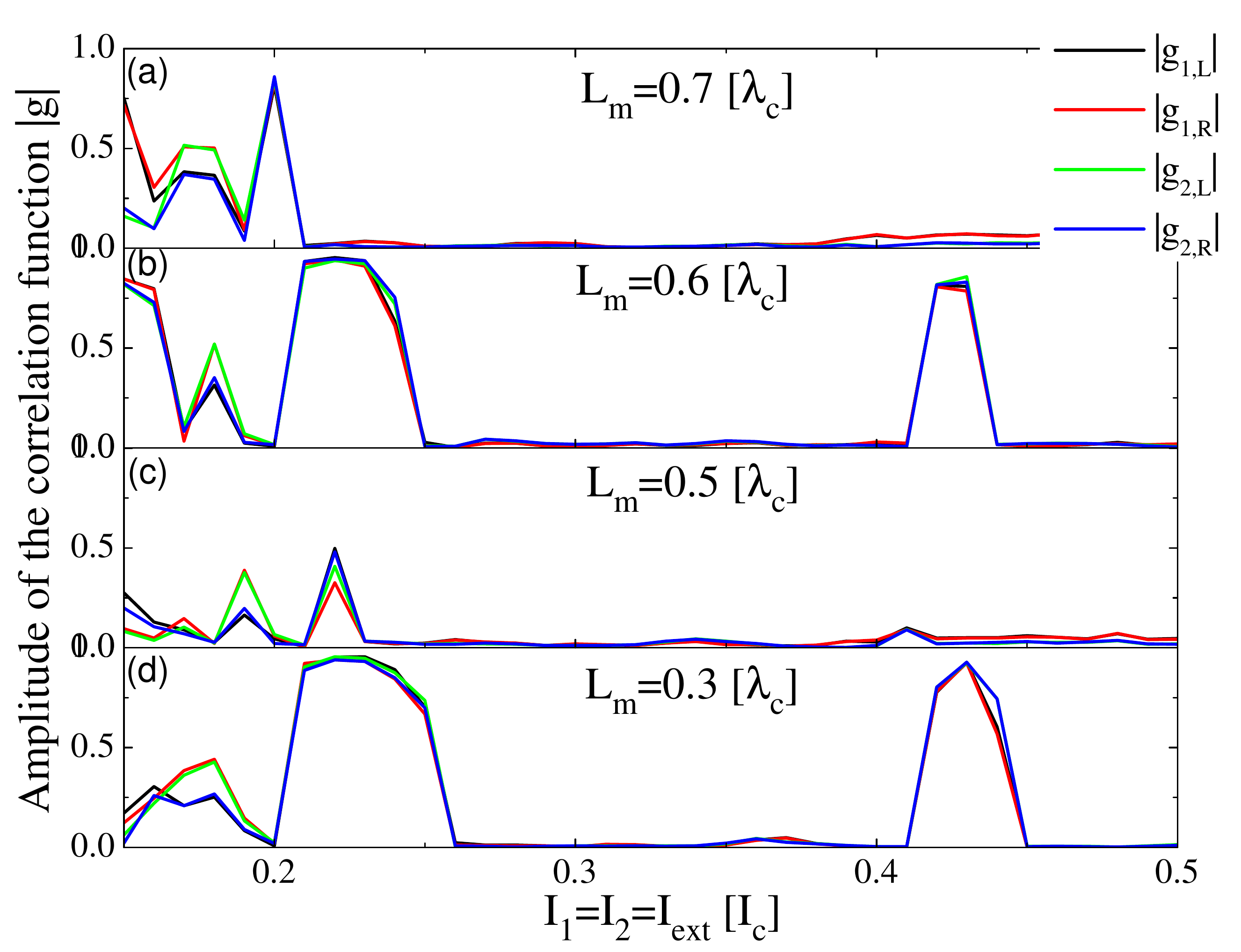,width=\columnwidth}
\caption{\label{f5}(color online) Phase coherence of plasma oscillations between
different stacks for different $L_m$. The phase coherence is measured by
$|g_{j,L/R}|$ defined in Eq. (\ref{eqg1}).}
\end{figure}

\begin{figure*}[t]
\psfig{figure=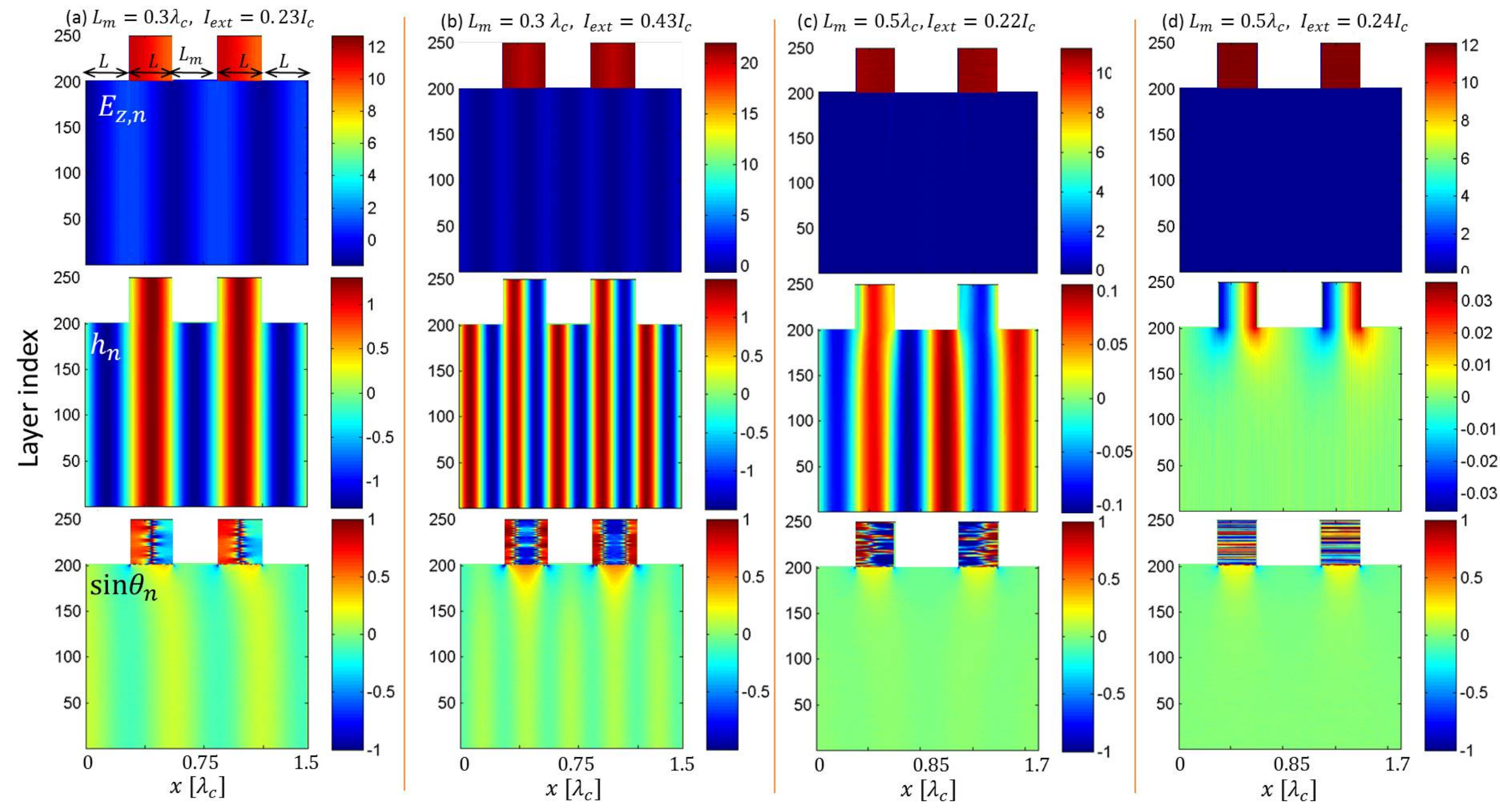,width=18cm} \caption{\label{f6}(color online) Snapshots of
the electric field (first row), magnetic field (second row) and Josephson
current $\sin (\theta_n)$ (third row) in the whole system. (a) and (b) are
obtained at the first and second cavity resonance for $L_m=L$. (c) and (d)
are results near the first cavity resonance for $L_m\neq n L$.}
\end{figure*}

\section{Numerical Simulations}

To find the condition for synchronization between mesas and checked the
stability of the synchronized state, we solve Eqs. (\ref{eq1}) and (\ref{eq2})
numerically for two mesas and numerical details are presented in Ref.
\onlinecite{szlin12a}. The number of junctions in the base is $N_c=200$ and in
the mesa is $N_m=50$. We take $\nu_{c}=0.02$, $\nu_{ab}=0.2$ and
$\ell\!=\!266.5$. To ensure that the resulting state is
stable, we add an artificial weak white noise current in Eq. (\ref{eq1}) in simulations,
$\left\langle \tilde{I}_{n}(x, t)\tilde{I}_{n'}(x',
t')\right\rangle=10^{-5}\delta(x-x')\delta(t-t')\delta(n-n')$.  To study the
coherence between different stack, we introduce an order parameter at edges of
mesa
\begin{equation}\label{eqr1}
r_{\mu,L/R}=\frac{1}{N_m}\sum_n^{N_m} \exp(i \theta_{n, L/R}^{\mu}),
\end{equation}
where $\theta_{n, L/R}^{\mu}$ is the phase difference of $n$-th layer at the left (L) or right (R)
edge of the $\mu$-th mesa. The time average of $|r_{\mu, L/R}|$, $\bar{r}_{\mu,
L/R}=\int_0^T |r_{\mu, L/R}| dt /T$ measures the phase coherence at the edges of
the mesa. For coherent oscillations of phase difference $r_{\mu,L/R}=1$ and for
completely random oscillations $r_{\mu,L/R}\rightarrow 0$ when $N_m\rightarrow \infty$.
To quantify the phase coherence between different mesas, we introduce a
correlation function
\begin{equation}\label{eqg1}
g_{\mu,L/R}=\frac{1}{T}\int_0^T r_{1,L}^* r_{\mu, L/R} dt,
\end{equation}
where we have taken the left edge of the first mesa as reference. Similarly
$|g_{\mu,L/R}|$ measures the coherence between the phase at left or right edges
of the $\mu$-th mesa and the phase at the left edge of the first mesa, and the
phase of $g_{\mu,L/R}$ represents the phase shift between them.

 \begin{figure}[b]
\psfig{figure=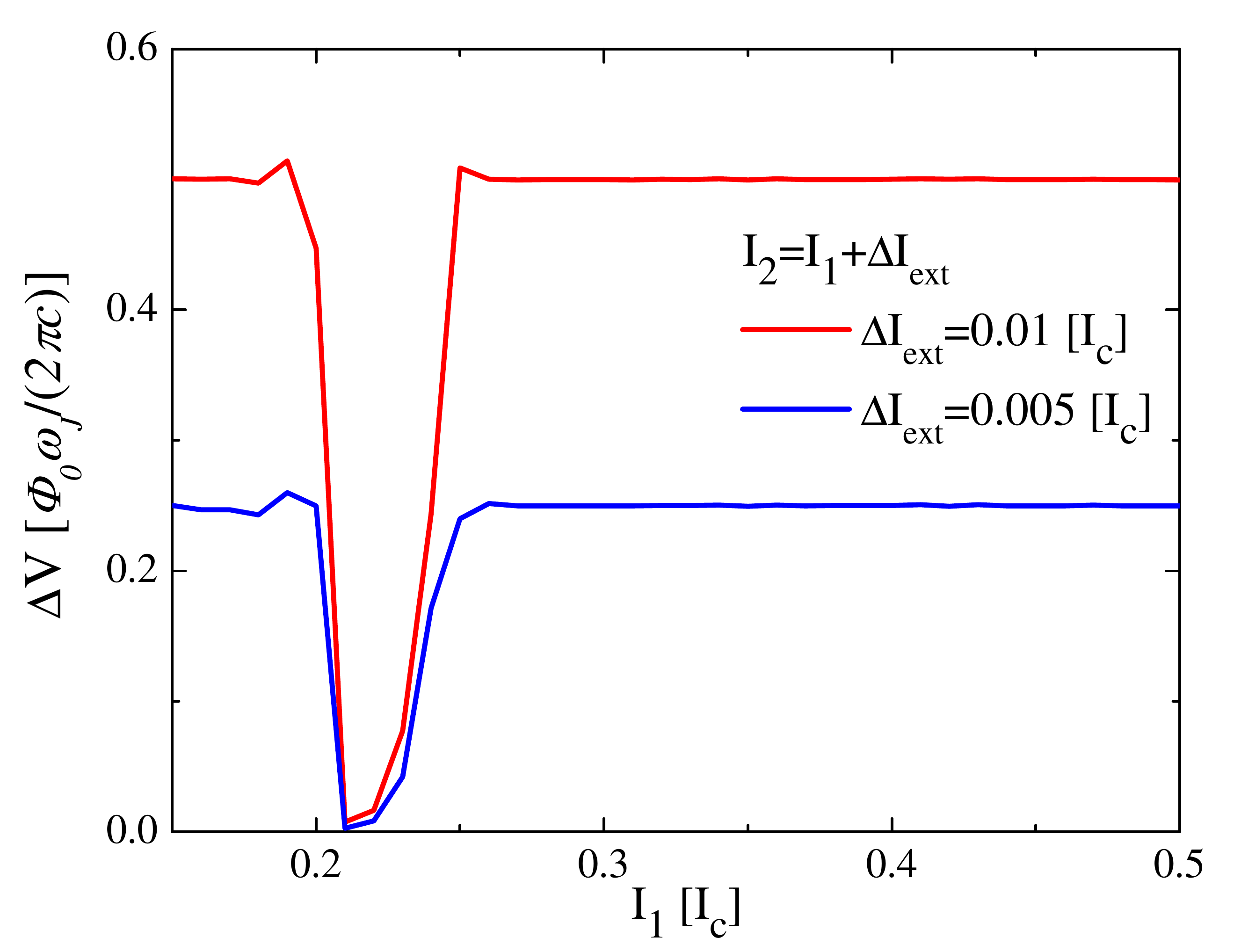,width=\columnwidth}
\caption{\label{f7}(color online) Voltage difference between two mesas $\Delta
V=V_2-V_1$ when the mesas are biased by different current $I_2=I_1+\Delta
I_{\rm{ext}}$.}
\end{figure}

Let us first consider two identical mesas with width $L=0.3\lambda_c$ and
with a separation $L_m$. They locate at the position $L$ away from the edges
of the base crystal. They are biased by the same current, $I_1=I_2=I_{\rm{ext}}$,
as shown in Fig. \ref{f6}(a). The results for different $L_m$ is shown in Fig.
\ref{f5} . The main peak at the left side corresponds to the fundamental cavity
mode of the mesa $\omega_{1}=\ell\pi/L$ and the the peak at the right side
corresponds to the second cavity mode $\omega_2=2\ell\pi/L$. When the frequency
of the plasma oscillations in the mesa matches the cavity frequency $\omega_m=m
\ell\pi/L$, $|g_{j,L/R}|$ increases indicating a tendency of synchronization
between two stacks. When $L_m= n L$, the phase coherence
between different mesas becomes maximal. This becomes clearer for the second
cavity mode, where two stacks do not synchronize at all when $L_m\neq n L$. For $L_m=0.7\lambda_c$, the peak at $I_{\rm{ext}}=2.0I_c$ is due to the cavity resonance inside the mesa. However the resonance occurs at smaller $I_{\rm{ext}}$ compared to that with $L_m = n L$. The downshift is due to the strong radiation damping through the base crystal as shown in Fig. \ref{f3}.

The reasons for the better coherence when $L_m= n L$  are as follows. For the
plasma oscillations uniform along the $c$ axis $q_k=0$, the in-plane dissipation
is absent and the plasma is damped by the weak dissipation
along the $c$-axis according to Eq. (\ref{qm}). However, for nonuniform oscillations with a finite wavenumber
$q_k$, the in-plane dissipation becomes
dominant for $N_m\approx 10^3$, and the nonuniform plasma oscillations in the base crystal decays
quickly. Therefore the interaction between two mesas is weak and the
synchronization becomes difficult. When $L_m=n L$, uniform cavity modes
$q_k=0$ are possible as shown in Fig. \ref{f6} (a) and (b). Two mesas then are
strongly coupled through the base crystal and they are synchronized. When $L_m\neq
n L$, only nonuniform modes can be excited and the plasma oscillations in the
base crystal is strongly damped by the in-plane dissipation. The amplitude of
plasma oscillations is small compared to that when $L_m=n L$, see Fig.
\ref{f6}(c) and (d). Thus synchronization between mesas is hard to attain.
Therefore the maximal synchronization is achieved when the position and size of
mesas are commensurate with the standing wave in the base crystal, because the
nonuniform plasma oscillations decay quickly in the base crystal.

Let us consider the phase shift of the gauge invariant phase difference between
edges of mesas. As shown in Fig. \ref{f6}(a), the supercurrent changes sign
from the left edge to the right edge in the same mesa. This indicates there is
a $\pi$ phase jump at the center of the mesa, and the $\pi$ phase kink is
excited at the cavity resonance\cite{szlin08b,Koshelev08b}. The $\pi$ phase
kink helps to pump energy into the plasma oscillations and the amplitude of the
oscillations is enhanced sharply at the cavity resonances, as described by the term at the right-hand side of Eq. (\ref{EqMesaPh}). In Fig. \ref{f6}(a) and (b), the
plasma oscillations at the left/right edges have the same phase between
different stacks when $L_m=(2n+1)L$. For $L_m=2n L$, there is $\pi$ phase shift between two mesas to match the standing wave in the base crystal.

In Fig. \ref{f7}, the voltage of mesas with $L_m=L$ when they are biased by
different current $I_{2}=I_{1}+\Delta I_{\rm{Iext}}$ is shown. At the cavity
resonance when two mesas are synchronized, they have the same voltage despite
that they are biased by slightly different current. Away from the cavity
resonance, two mesas decouple from each other and they oscillate at different
frequency.

\section{Conclusions}

We have investigated the synchronization of mesa array through the plasma oscillations in the base crystal. If one regards the mesa arrays and base crystal as a whole, the plasma oscillations inside depends on the configuration of mesa arrays as a result of geometrical resonance. The amplitude of the plasma oscillations and the tendency of synchronization is governed by the dissipation of the whole system, hence is determined by the configuration of mesa array. When the period of mesa array $a$ close to the multiple integer of the mesa width $L$, $a\approx n L$, the dissipation is minimized and mesas are synchronized at cavity resonances of the whole system. Alternatively, one may treat mesa and base crystal separately. When the cavity resonance of mesa matches of that in the base crystal, mesa array is synchronized. Otherwise, the cavity resonance of the mesas is suppressed by the strong dissipation due to the radiation into the base crystal, and the mesa array is not synchronized. Therefore the optimal configuration for synchronization is $a\approx n L$. The above picture is corroborated by both analytical calculations and numerical simulations.

\vspace{2mm}

 \noindent {\it Acknowledgements --}The authors thanks T. M. Benseman, U. Welp,
 and L. N. Bulaevskii for helpful discussions. SZL gratefully acknowledges
 funding support from the Office of Naval Research via the Applied
Electrodynamics collaboration. AEK is supported by UChicago Argonne, LLC,
operator of Argonne National Laboratory, a US DOE laboratory, operated under
contract No. DE-AC02-06CH11357.

%

\end{document}